\documentstyle[twoside,psfig]{article}

\catcode`\@=11
\long\def\@makefntext#1{
\protect\noindent \hbox to 3.2pt {\hskip-.9pt
$^{{\eightrm\@thefnmark}}$\hfil}#1\hfill}		

\def\@makefnmark{\hbox to 0pt{$^{\@thefnmark}$\hss}}	

\def\ps@myheadings{\let\@mkboth\@gobbletwo
\def\@oddhead{\hbox{}
\rightmark\hfil\eightrm\thepage}
\def\@oddfoot{}\def\@evenhead{\eightrm\thepage\hfil
\leftmark\hbox{}}\def\@evenfoot{}
\def\sectionmark##1{}\def\subsectionmark##1{}}

\oddsidemargin=\evensidemargin
\newcounter{sectionc}\newcounter{subsectionc}\newcounter{subsubsectionc}
\renewcommand{\section}[1] {\vspace{12pt}\addtocounter{sectionc}{1}
\setcounter{subsectionc}{0}\setcounter{subsubsectionc}{0}\noindent
	{\tenbf\thesectionc. #1}\par\vspace{5pt}}
\renewcommand{\subsection}[1] {\vspace{12pt}\addtocounter{subsectionc}{1}
	\setcounter{subsubsectionc}{0}\noindent
	{\bf\thesectionc.\thesubsectionc. {\kern1pt \bfit #1}}\par\vspace{5pt}}
\renewcommand{\subsubsection}[1] {\vspace{12pt}\addtocounter{subsubsectionc}{1}
	\noindent{\tenrm\thesectionc.\thesubsectionc.\thesubsubsectionc.
	{\kern1pt \tenit #1}}\par\vspace{5pt}}
\newcommand{\nonumsection}[1] {\vspace{12pt}\noindent{\tenbf #1}
	\par\vspace{5pt}}

\newcounter{appendixc}
\newcounter{subappendixc}[appendixc]
\newcounter{subsubappendixc}[subappendixc]
\renewcommand{\thesubappendixc}{\Alph{appendixc}.\arabic{subappendixc}}
\renewcommand{\thesubsubappendixc}
	{\Alph{appendixc}.\arabic{subappendixc}.\arabic{subsubappendixc}}

\renewcommand{\appendix}[1] {\vspace{12pt}
        \refstepcounter{appendixc}
        \setcounter{figure}{0}
        \setcounter{table}{0}
        \setcounter{lemma}{0}
        \setcounter{theorem}{0}
        \setcounter{corollary}{0}
        \setcounter{definition}{0}
        \setcounter{equation}{0}
        \renewcommand{\thefigure}{\Alph{appendixc}.\arabic{figure}}
        \renewcommand{\thetable}{\Alph{appendixc}.\arabic{table}}
        \renewcommand{\theappendixc}{\Alph{appendixc}}
        \renewcommand{\thelemma}{\Alph{appendixc}.\arabic{lemma}}
        \renewcommand{\thetheorem}{\Alph{appendixc}.\arabic{theorem}}
        \renewcommand{\thedefinition}{\Alph{appendixc}.\arabic{definition}}
        \renewcommand{\thecorollary}{\Alph{appendixc}.\arabic{corollary}}
        \renewcommand{\theequation}{\Alph{appendixc}.\arabic{equation}}
        \noindent{\tenbf Appendix \theappendixc #1}\par\vspace{5pt}}
\newcommand{\subappendix}[1] {\vspace{12pt}
        \refstepcounter{subappendixc}
        \noindent{\bf Appendix \thesubappendixc. {\kern1pt \bfit #1}}
	\par\vspace{5pt}}
\newcommand{\subsubappendix}[1] {\vspace{12pt}
        \refstepcounter{subsubappendixc}
        \noindent{\rm Appendix \thesubsubappendixc. {\kern1pt \tenit #1}}
	\par\vspace{5pt}}

\newcommand{\textlineskip}{\baselineskip=13pt}
\newcommand{\smalllineskip}{\baselineskip=10pt}

\def\eightcirc{
\begin{picture}(0,0)
\put(4.4,1.8){\circle{6.5}}
\end{picture}}
\def\eightcopyright{\eightcirc\kern2.7pt\hbox{\eightrm c}}

\newcommand{\copyrightheading}[1]
	{\vspace*{-2.5cm}\smalllineskip{\flushleft
	{\footnotesize Modern Physics Letters A, #1}\\
	{\footnotesize $\eightcopyright$\, World Scientific Publishing
	 Company}\\
	 }}

\newcommand{\publisher}[2]{{\begin{center}\footnotesize\smalllineskip
	Received #1\\
	Revised #2
	\end{center}
	}}

\def\abstracts#1#2#3{{
	\centering{\begin{minipage}{4.5in}\footnotesize\baselineskip=10pt
	\parindent=0pt #1\par
	\parindent=15pt #2\par
	\parindent=15pt #3
	\end{minipage}}\par}}

\def\keywords#1{{
	\centering{\begin{minipage}{4.5in}\footnotesize\baselineskip=10pt
	{\footnotesize\it Keywords}\/: #1
	 \end{minipage}}\par}}

\newcommand{\bibit}{\nineit}
\newcommand{\bibbf}{\ninebf}
\renewenvironment{thebibliography}[1]
	{\frenchspacing
	 \ninerm\baselineskip=11pt
	 \begin{list}{\arabic{enumi}.}
        {\usecounter{enumi}\setlength{\parsep}{0pt}
	 \setlength{\leftmargin 12.7pt}{\rightmargin 0pt} 
         \setlength{\itemsep}{0pt} \settowidth
	{\labelwidth}{#1.}\sloppy}}{\end{list}}

\newcounter{itemlistc}
\newcounter{romanlistc}
\newcounter{alphlistc}
\newcounter{arabiclistc}

\newcommand{\fcaption}[1]{
        \refstepcounter{figure}
        \setbox\@tempboxa = \hbox{\footnotesize Fig.~\thefigure. #1}
        \ifdim \wd\@tempboxa > 5in
           {\begin{center}
        \parbox{5in}{\footnotesize\smalllineskip Fig.~\thefigure. #1}
            \end{center}}
        \else
             {\begin{center}
             {\footnotesize Fig.~\thefigure. #1}
              \end{center}}
        \fi}

\newcommand{\tcaption}[1]{
        \refstepcounter{table}
        \setbox\@tempboxa = \hbox{\footnotesize Table~\thetable. #1}
        \ifdim \wd\@tempboxa > 5in
           {\begin{center}
        \parbox{5in}{\footnotesize\smalllineskip Table~\thetable. #1}
            \end{center}}
        \else
             {\begin{center}
             {\footnotesize Table~\thetable. #1}
              \end{center}}
        \fi}

\def\@citex[#1]#2{\if@filesw\immediate\write\@auxout
	{\string\citation{#2}}\fi
\def\@citea{}\@cite{\@for\@citeb:=#2\do
	{\@citea\def\@citea{,}\@ifundefined
	{b@\@citeb}{{\bf ?}\@warning
	{Citation `\@citeb' on page \thepage \space undefined}}
	{\csname b@\@citeb\endcsname}}}{#1}}

\newif\if@cghi
\def\cite{\@cghitrue\@ifnextchar [{\@tempswatrue
	\@citex}{\@tempswafalse\@citex[]}}
\def\citelow{\@cghifalse\@ifnextchar [{\@tempswatrue
	\@citex}{\@tempswafalse\@citex[]}}
\def\@cite#1#2{{$\null^{#1}$\if@tempswa\typeout
	{IJCGA warning: optional citation argument
	ignored: `#2'} \fi}}

\def\pmb#1{\setbox0=\hbox{#1}
	\kern-.025em\copy0\kern-\wd0
	\kern.05em\copy0\kern-\wd0
	\kern-.025em\raise.0433em\box0}

\def\fnt#1#2{\footnotetext{\kern-.3em
	{$^{\mbox{\scriptsize #1}}$}{#2}}}

\def\fpage#1{\begingroup
\voffset=.3in
\thispagestyle{empty}\begin{table}[b]\centerline{\footnotesize #1}
	\end{table}\endgroup}

\def\runninghead#1#2{\pagestyle{myheadings}
\markboth{{\protect\footnotesize\it{\quad #1}}\hfill}
{\hfill{\protect\footnotesize\it{#2\quad}}}}
\font\tenrm=cmr10
\font\tenit=cmti10
\font\tenbf=cmbx10
\font\bfit=cmbxti10 at 10pt
\font\ninerm=cmr9
\font\nineit=cmti9
\font\ninebf=cmbx9
\font\eightrm=cmr8

\def\qed{\hbox{${\vcenter{\vbox{			
   \hrule height 0.4pt\hbox{\vrule width 0.4pt height 6pt
   \kern5pt\vrule width 0.4pt}\hrule height 0.4pt}}}$}}

\begin{document}
\setlength{\textheight}{7.7truein}  

\runninghead{D. V. Ahluwalia and M. Kirchbach}
{$(1/2,1/2)$ Representation space $\ldots$}

\normalsize\textlineskip
\thispagestyle{empty}
\setcounter{page}{1}

\copyrightheading{Vol. 16, No.1 (2001) 1377--1383}

\vspace*{0.88truein}

\fpage{1}
\centerline{\bf $\mathbf
(1/2,1/2)$ REPRESENTATION SPACE}
\baselineskip=13pt
\centerline{\bf ---  AN AB INITIO CONSTRUCT ---}

\vspace*{0.37truein}
\centerline{\footnotesize D. V. AHLUWALIA\footnote{Summer address:
Group P-25, Mail Stop
H-846, Los Alamos
National Laboratory, Los Alamos, NM 87545, USA.
E-mail: ahluwalia@phases.reduaz.mx;
http://phases.reduaz.mx}~  and
M. KIRCHBACH\footnote{
E-mail: kirchbach@chiral.reduaz.mx; http://chiral.reduaz.mx} }
\baselineskip=12pt
\centerline{\footnotesize\it
Theoretical Physics Group}
\baselineskip=10pt
\centerline{\footnotesize\it
Facultad de Fisica, Univ. Aut. de Zacatecas}
\baselineskip=10pt
\centerline{\footnotesize\it ISGBG, Ap. Postal C-600, Zacatecas,
ZAC 98062, Mexico}

\vspace*{10pt}

\publisher{(received date)}{(revised date)}

\vspace*{0.21truein}
\abstracts{A careful {\em ab initio\/} construction of the
finite-mass $(1/2,1/2)$
representation space of the Lorentz group reveals it to be a spin-parity
multiplet. In general, it does not lend itself to a single-spin
interpretation. We find that the $(1/2,1/2)$ representation space for massive
particles naturally bifurcates into a triplet and a singlet of opposite
relative intrinsic parties. The text-book separation into spin one  and
spin zero states occurs only for certain limited kinematical settings.
We construct a wave equation for the  $(1/2,1/2)$ multiplet,
and show that the particles and antiparticles in this representation
space do not carry a definite spin but only a definite relative intrinsic
parity. In general, both spin one and spin zero are
covariantly inseparable inhabitants  of massive vector fields. This last
observation suggests that scalar particles, such as the Higgs,
are natural inhabitants of massive  $(1/2,1/2)$
representation space.}{}{}

\vspace*{10pt}

\textlineskip			
\vspace*{12pt}			

{\sc Spin-Parity Lorentz Multiplets}
complement the supersymmetry multiplets.
The former are characterized by purely bosonic (or fermionic)
constituents, while the latter carry bosonic and fermionic
degrees of freedom simultaneously.
To establish this assertion, via an example,  we here
present an {\em ab initio} study of the $(1/2,1/2)$ representation
space and find that it is much richer than usually believed. The new
physical element arises from fully respecting the algebra of the
Lorentz group. Specifically, since the generators of rotations and
boosts do not commute in general, we will find that  a massive
$(1/2,1/2)$ representation space supports an
irreducible multiplet of spin 1 and spin 0. The separability
of spin 1, and spin 0, occurs only in some kinematically restricted
settings.

The construction of the $(1/2,1/2)$ representation
space begins with the observation that it is a direct
product of the $(1/2,0)$ and $(0,1/2)$ representation spaces, i.e.,
 $(1/2,1/2) = (1/2,0)\otimes(0,1/2)$. Under the Lorentz boost
the right-handed  $(1/2,0)$ spinors, $\phi_R(\vec{p\, })$,
transform as
\def\beq{\begin{eqnarray}}
\def\eeq{\end{eqnarray}}

\beq
\phi_R\left(\vec p\,\right)=
\exp\left(+\, \frac{\vec \sigma}{2}\cdot\vec\varphi\right)\phi_R
(\vec 0\,)=
\frac{1}{\sqrt{2 m (E+m)}}\Big[
(E+m)I_2+\vec\sigma\cdot\vec p\,\Big] \phi_R
(\vec 0\,)\,,
\eeq
where $\vec\varphi$ is the boost parameter,\cite{lhr} and $I_2$ is a
$2\times 2$ identity matrix. The
left-handed $(0,1/2)$ spinors, $\phi_L(\vec{p\, })$, transform
according to
\beq
\phi_L(\vec p\,)=
\exp\left(-\,\frac{\vec \sigma}{2}\cdot\vec\varphi\right)\phi_L
(\vec 0\,)=
\frac{1}{\sqrt{2 m (E+m)}}\Big[
(E+m)I_2 - \vec\sigma\cdot\vec p\,\Big] \phi_L
(\vec 0\,)\,.
\eeq
Now to describe the physical states inhabiting the $(1/2,1/2)$
representation space, one needs four independent {\em rest}
states, and  the {\em boost}, $\kappa(\vec{p\, })$, given by
\beq
\kappa(\vec p\,)=\frac{1}{2 m (E+m)}\Big[
(E+m)I_2+\vec\sigma\cdot\vec p\,\Big]\otimes
\Big[
(E+m)I_2-\vec\sigma\cdot\vec p\,\Big]\,.
\eeq
Associated with $\kappa(\vec p\,)$ are the boost generators
\beq
K_x=
\frac{1}{2}\left[
\begin{array}{cccc}
0 & i & -i & 0 \\
i & 0 & 0 & -i \\
-i & 0 & 0 & i \\
0 & -i & i & 0
\end{array}
\right],\quad
K_y=
\frac{1}{2}\left[
\begin{array}{cccc}
0 & 1 & -1 & 0\\
-1 & 0 & 0 & -1 \\
1 & 0 & 0 & 1 \\
0 & 1 & -1 & 0
\end{array}
\right],
\eeq
\beq
K_z=
\left[
\begin{array}{cccc}
0 & 0 & 0 & 0 \\
0 & -i & 0 & 0 \\
0 & 0 & i & 0\\
0 & 0 & 0 & 0
\end{array}
\right]\,.
\eeq
Similarly, the generators of the rotations for the
$(1/2,1/2)$ representation space are:
\beq
J_x=
\frac{1}{2}\left[
\begin{array}{cccc}
0&1&1&0\\
1&0&0&1\\
1&0&0&1\\
0&1&1&0
\end{array}
\right],\quad
J_y=
\frac{1}{2}\left[
\begin{array}{cccc}
0&-i&-i&0\\
i&0&0&-i\\
i&0&0&-i\\
0&i&i&0
\end{array}
\right],
\eeq
\beq
J_z=
\left[
\begin{array}{cccc}
1&0&0&0\\
0&0&0&0\\
0&0&0&0\\
0&0&0&-1
\end{array}
\right]\,.
\eeq
{\bf In the  rest frame},
the $(1/2,1/2)$ representation
space can be decomposed into eigenstates of
$\vec J^{\,2}$ and  $J_z$. The spin-$1$ sector
carries three independent rest-frame states:
\beq
w_{1,+1}(\vec 0\,)=
\left[
\begin{array}{c}
1\\
0\\
0\\
0\\
\end{array}
\right],\,\,
w_{1,0}(\vec 0\,)=
\frac{1}{\sqrt{2}}\left[
\begin{array}{c}
0\\
1\\
1\\
0\\
\end{array}
\right]
,\,\,
w_{1,-1}(\vec 0\,)=\left[
\begin{array}{c}
0\\
0\\
0\\
1\\
\end{array}
\right],
\eeq
while the spin-$0$ sector carries one state only
\beq
w_{0,0}(\vec 0\,)=
\frac{1}{\sqrt{2}}\left[
\begin{array}{c}
0\\
1\\
-1\\
0\\
\end{array}
\right].
\eeq
{\bf By the Wigner argument}, $\kappa (\vec p\,)\,w_{j,m}
(\vec 0\,)$  can be identified
with the states corresponding to momentum $\vec p$.
The explicit expressions for  these are:
\beq
w_1(\vec p\,):= \kappa(\vec p\,) w_{1,+1}(\vec 0\,)={N}
\left[
\begin{array}{c}
(E+m-p_z)(E+m+p_z)\\
-(p_x+i p_y)(E+m+p_z)\\
(p_x+ i p_y)(E+m-p_z)\\
-(p_x+i p_y)^2
\end{array}
\right]\,,
\eeq

\beq
w_2(\vec p\,):=\kappa(\vec p\,) w_{1,0}(\vec 0\,)=\frac{N}{\sqrt{2}}
\left[
\begin{array}{c}
-2 (p_x- i p_y) p_z\\
\left(p_z^2+ 2\, m\, p_z + 2\, E\, p_z - p_y^2 -p_x^2 + (E+m)^2 \right) \\
\left(p_z^2- 2\, m\, p_z - 2\, E\, p_z - p_y^2 -p_x^2 + (E+m)^2 \right) \\
2 (p_x+i p_y)p_z
\end{array}
\right]\,,
\eeq

\beq
w_3(\vec p\,) := \kappa(\vec p\,) w_{1,-1}(\vec 0\,)={N}
\left[
\begin{array}{c}
- (p_x - i p_y)^2 \\
(p_x-i p_y)(E+m+p_z)\\
- (p_x- i p_y)(E+m-p_z) \\
(E+m-p_z)(E+m+p_z)
\end{array}
\right]\,,
\eeq

\beq
w_4(\vec p\,):=\kappa(\vec p\,) w_{0,0}(\vec 0\,)
=\frac{N}{\sqrt{2} }
\left[
\begin{array}{c}
-2 (p_x- i p_y) (E+m)\\
\left(p_z^2+ 2\, m\, p_z + 2\, E\, p_z + p_y^2 +p_x^2 + (E+m)^2 \right) \\
- \left(p_z^2- 2\, m\, p_z - 2\, E\, p_z + p_y^2 +p_x^2 + (E+m)^2 \right) \\
2 (p_x+i p_y)(E+m)
\end{array}
\right]\,,
\eeq
where the normalization factor takes the value,
\beq
N := \left[{2 m (E+m)}\right]^{-1}.
\eeq
The
new labeling
$w_\zeta(\vec p\,)$, $\zeta=1,2,3,4$,
has been introduced to emphasize that, in general,
$w_\zeta(\vec p\,)$ are {\em not\/} single-spin valued states.
Since the generators of rotations commute with the generators of
boost only in case the rotation plane is perpendicular
to the direction of motion, a general decomposition of the (1/2,1/2)
representation space into the spin 1 and spin 0 sectors is ruled out.
Stated differently, since $\vec J^{\,2}$ is not a Casimir
operator of the
Poincar\'e group, a generally valid decomposition of
the $(1/2,1/2)$ representation space into spin 1 and spin 0
sectors is forbidden.

The exception occurs for the helicity basis  in which one
chooses the physical setting to be such that the ``quantization
axis'' for the spin-projections is aligned along the direction
of motion. More specifically,
$[J_\imath,K_\jmath] $
vanish only if $\imath$ equals $\jmath$.
For instance, by setting, $p_x=0=p_y$, in the above expressions
for $w_\zeta(\vec p\,)$ in order to produce a special case of the
helicity basis, we find that
(a) The $w_1(\vec p\,)$- and  $w_3(\vec p\,)$-states are
eigenvectors of $\vec J^{\,2}$ and $J_z$ with eigenvalues $2 [=1(1+1)]$,
$\pm 1$, respectively; and
(b) A third spin-1 state  -- to be denoted by $v_2(\vec{p\, }) $ --
can be obtained from either $w_3(\vec{p\, } )$, or $w_1(\vec{p\, })$,
through the action of the ladder operators
$J_+:=J_x+i J_y$, and $J_-:=J_x-i J_y$,
respectively. The state $v_2(\vec{p\, }):=
\left(1/\sqrt{2}\right)J_{+} \,w_3(\vec{p\, } ) =
\left(1/\sqrt{2}\right)J_{-} \,w_1(\vec{p\, } )$
turns out to be
simultaneous eigenvector of $\vec J^{\,2}$ and $J_z$ with eigenvalues
$2  [=1(1+1)]$ and $0$, respectively.
Note, that $v_2(\vec{p\, })$ is essentially different from
$w_2(\vec p\,)$.
The three states $\{w_1(\vec{p\, }),
v_2(\vec{p\, }), w_3(\vec{p\, })\}$ constitute a
spin-$1$ multiplet;
(c) Finally, a fourth state $v_4(\vec p\,)$,
orthonormal to the previous ones,
can be constructed and
shown to be a  genuine spin-$0$ state.
Thus, the particular cases of the helicity basis from above,
on the one hand, and the rest frame, on the other hand,
represent two kinematically-restricted scenarios
where the $(1/2,1/2)$ space can split into spin $0$, and
spin $1$.

The orthonormality relations for  $w_\zeta(\vec p\,)$ are
\beq
\overline{w}_\zeta(\vec p\,)\, w_\zeta(\vec p\,)
 =\cases{-1\,\, \mbox{for}\,\, \zeta=1,2,3 \cr
                                    +1\,\, \mbox{for}\,\, \zeta=4}
\eeq
while the completeness relation reads
\beq
w_4(\vec p\,)\,\overline{w}_4(\vec p\,)
 - \sum_{\zeta=1}^3 w_\zeta(\vec p\,)\,\overline{w}_\zeta(\vec p\,)
=I_4\,,\label{cr}
\eeq
where $I_4$ equals $4\times 4$ identity matrix.  In the above expressions
we have defined
\beq
\overline{w}_\zeta(\vec p\,):=w_\zeta(\vec p\,)^\dagger\,\lambda_{00},
\quad
\mbox{where} \quad
\lambda_{00}=
\left[
\begin{array}{cccc}
-1&0&0&0\\
0&0&-1&0\\
0&-1&0&0\\
0&0&0&-1
\end{array}
\right]\,.
\eeq
It will be seen below that $\lambda_{00}$
is a part of a larger set of covariant matrices. For the moment
it suffices to note that $S \lambda_{00} S^{-1}$, with
\beq
S=
\frac{1}{\sqrt{2}}\left[
\begin{array}{cccc}
0&i&-i&0\\
-i&0&0&i\\
1&0&0&1\\
0&i&i&0
\end{array}
\right]\,,
\eeq
is the flat space-time metric with the diagonal  $\{1,-1,-1,-1\}$.
We note that $ \lambda_{00}$ also serves as the {\em Parity\/} operator:
\beq
\lambda_{00}\, w_\zeta(-\vec p\,)
 =\cases{-\, w_\zeta(\vec p\,)\,\,\mbox{for}\,\,  \zeta=1,2,3 \cr
         + \,w_\zeta(\vec p\,) \,\,    \mbox{     for}\,\,   \zeta=4\cr}
\eeq
So, while $w_ \zeta(\vec p\,)$ may not carry a definite spin they
are endowed with a definite  relative intrinsic parity. The relevant
projectors onto sub-spaces of opposite relative intrinsic parities are:
\beq
P_{+}= w_4(\vec p\,)\,\overline{w}_4(\vec p\,), \,\,\mbox{and}\,\,
P_{-}= -\, \sum_{ \zeta=1}^3 w_\zeta(\vec p\,)\,
\overline{w}_ \zeta(\vec p\,)\,.\label{proj}
\eeq

We now note that $S  w_\zeta(\vec p\,)$ transforms as a Lorentz four vector
and carries a Lorentz index that can be lowered and raised with the
flat space-time metric pointed out above. Thus, we introduce:
\beq
A^\mu_\zeta(\vec p\,):=S  w_\zeta(\vec p\,)\,,
\eeq
and, additionally
\beq
{\cal P}_{\pm} := S P_{\pm} S^{-1},
\eeq
and
\beq
\Lambda_{\mu\nu} := S \lambda_{\mu\nu} S^{-1}.
\eeq
In a way reminiscent of the Dirac case,
the wave equation for $ A_\zeta(\vec p\,)$ can be read off from the
projectors and is seen to be:
\beq
\left(\Lambda_{\mu\nu}p^\mu p^\nu - \epsilon \,m^2 I_4\right)
A_\zeta(\vec p\,) = 0\,,\label{new}
\eeq
where $\epsilon$ equals $+1$ for $\zeta=4$ and is $-1$ for
$\zeta=1,2,3$. The $\Lambda_{\mu\nu}$ matrices can be read off
from
\beq
\Lambda_{\mu\nu}p^\mu p^\nu = m^2\left({\cal P}_{+}
- {\cal P}_{-}\right)\,.
\eeq
By studying
${\mbox{det}}\left(\Lambda_{\mu\nu}p^\mu p^\nu - \epsilon \,m^2 I_4\right)$,
we find that: (a)
For $\epsilon=-1$, the above equation carries three ``positive--energy--''
and three ``negative--energy--'' solutions with
the correct dispersion relation, $E^2=\vec p^{\,2} + m^2$, while (b) For
$\epsilon=+1$, there is one ``positive--energy--''and one
``negative--energy--'' solution.
We conjecture that a complete CPT analysis
would show these to be particle-antiparticle solutions.
Thus the $(1/2,1/2)$ representation space  manifestly
carries the particle and antiparticle states with equal masses.
This result is not contained in the usual considerations based
on the Proca equation. It is further apparent from the derived wave
equation that in the
$(1/2,1/2)$ representation space $[C,P]=0$. This contrasts with
$\{C,P\}=0$ for the $(1,0)\oplus(0,1)$ representation space [2,3].

The $a^\nu$ associated with the usual spin-1 Proca equation
\beq
\partial_\mu F^{\mu\nu} + m^2 a^\nu =0\,,\label{peq}
\eeq
by construction\footnote{$F^{\mu\nu}=\partial^\mu a^\nu
-\partial^\nu a^\mu$, is antisymmetric.}
satisfies $\partial_\nu a^\nu=0$ (for $m\ne 0$).
However, ``$\partial_\nu A^\nu=0$'' can not be implemented in the
massive $(1/2,1/2)$ representation space without violating the
completeness relation (\ref{cr}). While the new wave
equation (\ref{new}) contains all solutions of (\ref{peq}), the
converse is not true. For this reason, the Proca equation is not
endowed with the complete physical content of the  $(1/2,1/2)$
representation space. Its content is
confined entirely to the space associated with the
projector $P_-$ (see Eq. (\ref{proj})).

Every knowledgeable reader by now must have noted
as to how parallel is the presented construct with the Dirac
construct for spin one half, and that the usual ``completeness relation''
written in Proca framework is not a completeness relation in the usual
mathematical sense. The mathematically correct form of
the completeness relation is Eq.~(\ref{cr}). It is the exact parallel
of its counterpart
in the $(1/2,0)\oplus(0,1/2)$ Dirac construct.

In view of these observations,
we  now critically study
the ``constraint'':
\beq
\partial_\beta A^\beta = 0,
\eeq
which in the momentum-space takes the form,
\beq
p_\beta A^\beta_\zeta (\vec p\,) =0
\eeq
The latter ``constraint'' can be evaluated using the results
obtained above. We find
\beq
p_\beta A^\beta_\zeta (\vec p\,) =0 \
\Leftrightarrow
\cases{c_\zeta \left(m^2-p_\beta p^\beta\right)&=\,0,\,\,
\mbox{for}\,\, $\zeta=1,2,3$\cr
\qquad(i/m) p_\beta p^\beta &=\,0,\,\, \mbox{for}\,\, $\zeta=4$\cr}
\eeq
where,
$c_1=i(p_x+i p_y)$, $c_2=-ip_z$, and $c_3 -i(p_x-i p_y)$.
In other words,
the conditions for $\zeta=1,2,3$ yield $p_\beta p^\beta = m^2$,
which certainly is not an additional constraint for a Lorentz
covariant theory. For $\zeta =4$, $p_\beta A_4^\beta= i m$,
and cannot be set equal to zero unless one is considering massless
particles -- for which case it is  trivially satisfied
(and, again, does not constitute an additional
constraint).
We thus arrive at the conclusion that
$p_\beta A^\beta_\zeta(\vec p\,) =0$ is trivially satisfied for the
massless case. Therefore, it cannot be used as a ``constraint.'' On the
other hand, for the massive case it is equivalent to, $E^2=\vec p^{\,2}+m^2$,
for $\zeta=1,2,3$, whereas for $\zeta=4$ it can not
be fulfilled at all.

It is thus apparent from the above considerations that an isolated
kine\-mati\-cally-unrestricted massive spin-$1$ field theory can only be
based upon the $(1,0)\oplus(0,1)$ representation space.
However, contrary to the canonical wisdom,
the $(1,0)\oplus(0,1)$ representation space supports one of the unusual
Wigner classes.\cite{ew} In this representation space
massive bosons, and antibosons, carry {\em opposite} relative intrinsic
parities.\cite{bww} For this reason, e.g.,
the standard-model's $W^\pm$ and $Z$ bosons cannot
be described as $(1,0)\oplus(0,1)$ particles,
and we are left with  the $(1/2,1/2)$ representation space -- as is in fact
appropriate for gauge bosons.
On the other hand, the $(1/2,1/2)$ representation
space is a multiplet of spin $1$ and spin $0$. If single-spin valued
spin-$1$
particles are to be created in this representation space then,
(a) They must be spin-polarized along the direction of motion with
the eigenvalues of the helicity operator,
$\left({\vec p}/{\vert \vec p\vert}\right)\cdot \vec J$,
equal to $\{+1,0,-1\}$;
(b) The completeness relation for the $(1/2,1/2)$ representation space
obtained above requires  such
spin-$1$ particles to be necessarily accompanied by a spin-$0$ partner.
In the most general case, the $(1/2,1/2)$ representation space
naturally splits only into a triplet and a
singlet of opposite relative intrinsic parities, which are
not to
be confused with the spin-1 triplet and the spin-0 singlet.
This confirms an earlier conjecture by one of us\cite{mk} that
$(j,j)$, or in general $(j,j)\otimes
\lbrack(1/2,0)\oplus(0,1/2)\rbrack$, representation spaces do not carry
a single-spin interpretation. Instead, they should be interpreted as
spin-parity Lorentz multiplets, an observation empirically supported
by the existing data on the non-strange baryonic spectra.

\nonumsection{References}

\end{document}